\documentclass[usenatbib,onecolumn]{mnras}
\usepackage{epsfig}
\usepackage{amsmath,amssymb}
\usepackage{cancel}
\usepackage{graphicx}
\DeclareGraphicsExtensions{.png,.pdf}
\usepackage{txfonts}
\usepackage[english]{babel}
\usepackage[utf8]{inputenc}
\usepackage[T1]{fontenc}
\usepackage{fontenc}
\usepackage{soul}
\allowdisplaybreaks
\DeclareMathOperator\atanh{atanh}
\DeclareMathOperator\asinh{asinh}

\newcommand{\beq}{\begin{equation}}
\newcommand{\eeq}{\end{equation}} 

\newcommand\cs{c_{\rm s}}
\newcommand\csiso{c_{\rm s,iso}}

\newcommand\cszero{c_{\rm s,0}}

\renewcommand\d{{\rm d}}

\newcommand\de{\partial}
\newcommand\Deltaomega{\Delta_{\omega}}
\newcommand\Deltas{\Delta_{s}}
\newcommand\deltapress{\delta\press}
\newcommand\deltarho{\delta \rho}

\newcommand\Eq{Equation\ }
\newcommand\Eqs{Equations\ }

\newcommand\F{\mathcal{F}}

\newcommand\gammaprime{{\gamma^{\prime}}}

\newcommand\gammap{{\gammaprime}}
\newcommand\grad{{\nabla}}

\newcommand{\hz}{h_z}
\newcommand{\hXperc}{h_{X\%}}
\newcommand{\hsixty}{h_{60\%}}
\newcommand{\hseventy}{h_{70\%}}
\newcommand{\heighty}{h_{80\%}}
\newcommand{\imaginary}{{\rm i}}
\newcommand{\im}{\imaginary}
\renewcommand{\Im}{{\rm Im}}

\newcommand\kRinst{k_{R,{\rm inst}}}

\newcommand\kz{k_z}
\newcommand\kR{k_R}
\newcommand{\kJ}{k_{\rm J}}
\newcommand\kv{{\boldsymbol k}}

\renewcommand\L{\mathcal{L}}

\newcommand\N{\mathcal{N}}
\newcommand{\Nz}{N_z}

\newcommand\Phiext{\Phi_{\rm ext}}

\newcommand\press{p}

\newcommand\pressR{\press^{\prime}_R}
\newcommand\pressz{\press^{\prime}_z}
\newcommand\presszero{\press_0}

\newcommand{\Qthreed}{Q_{\rm 3D}}
\newcommand{\Qcrit}{Q_{\rm crit}}
\newcommand{\Qthreedmin}{Q_{\rm 3D,min}}
\newcommand{\qunp}{q_{\rm unp}}

\newcommand\rhobar{\bar{\rho}}

\newcommand\rhoR{\rho^{\prime}_R}
\newcommand\rhoz{\rho^{\prime}_z}

\newcommand\rhozero{\rho_0}

\newcommand\sech{{\rm sech}}

\newcommand\sigmaR{\sigma^{\prime}_R}
\newcommand\sigmaz{\sigma^{\prime}_z}

\newcommand{\uv}{{\boldsymbol u}}

\newcommand{\uphi}{u_\phi}

\newcommand{\uR}{u_R}

\newcommand{\uz}{u_z}

\newcommand\ztilde{\tilde{z}}

\newcommand\zxi{z_\xi}
\newcommand\zcritinst{z_{\rm crit,inst}}
\newcommand\zcritstab{z_{\rm crit,stab}}

\begin{document}

\date{November 16 2022}

\title[Gravitational instability of rotating fluids]{Local gravitational instability of stratified rotating fluids: 3D criteria for gaseous discs}

\author[C.\ Nipoti]{\parbox{\textwidth}{Carlo Nipoti$^{1}$\thanks{E-mail: carlo.nipoti@unibo.it}}\vspace{0.4cm}\\
  \parbox{\textwidth}{
    $^1$Dipartimento di Fisica e Astronomia ``Augusto Righi'', Alma Mater Studiorum -- Universit\`a di Bologna,
    via Gobetti 93/2, I-40129 Bologna, Italy}}

\maketitle 

\begin{abstract}
Fragmentation of rotating gaseous systems via gravitational
instability is believed to be a crucial mechanism in several
astrophysical processes, such as formation of planets in protostellar
discs, of molecular clouds in galactic discs, and of stars in
molecular clouds.  Gravitational instability is fairly well understood
for infinitesimally thin discs. However, the thin-disc approximation
is not justified in many cases, and it is of general interest to study
the gravitational instability of rotating fluids with different
degrees of rotation support and stratification.  We derive dispersion
relations for axisymmetric perturbations, which can be used to study
the local gravitational stability at any point of a rotating
axisymmetric gaseous system with either barotropic or baroclinic
distribution.  3D stability criteria are obtained, which generalize
previous results and can be used to determine whether and where a
rotating system of given 3D structure is prone to clump formation. For
a vertically stratified gaseous disc of thickness $\hz$ (defined as
containing $\approx$70\% of the mass per unit surface), a sufficient
condition for local gravitational instability is
$\Qthreed\equiv\left(\sqrt{\kappa^2+\nu^2}+\cs\hz^{-1}\right)/{\sqrt{4\pi
    G\rho}}<1$, where $\rho$ is the gas volume density, $\kappa$ the
epicycle frequency, $\cs$ the sound speed, and
$\nu^2\equiv\rhoz\pressz/\rho^2$, where $\rhoz$ and $\pressz$ are the
vertical gradients of, respectively, gas density and pressure. The
combined stabilizing effects of rotation ($\kappa^2$) and
stratification ($\nu^2$) are apparent. In unstable discs, the
conditions for instability are typically met close to the midplane,
where the perturbations that are expected to grow have characteristic
radial extent of a few $\hz$.
\end{abstract}


\section{Introduction}
\label{sec:intro}

Rotating gaseous structures confined by gravitational potentials are
widespread among astrophysical systems on a broad range of
scales. Prototypical examples include gaseous galactic discs,
accretion discs, and protostellar and protoplanetary discs, but
rotation can be non-negligible also in pressure-supported systems such
as stars, molecular clouds, galactic coronae and hot atmospheres of
galaxy clusters. The confining gravitational potential can be due
either only to the gas itself or to a combination of the gas
self-gravity and of an external potential. Whenever the gas
self-gravity is locally non-negligible with respect to the external
potential, the evolution of the rotating fluid depends crucially on
whether it is gravitationally stable or unstable. For instance, in
galactic discs local gravitational instability is expected to lead to
fragmentation, growth of dense gas clumps and eventually to star
formation \citep[see e.g.\ section 8.3 of][]{CFN19}.  In
protostellar discs, gravitational instability can contribute either
directly (via gas collapse) or indirectly (via concentration of dust
particles) to the process of planet formation \citep{Kra16}. It is
thus not surprising that the study of gravitational instability of
rotating fluids has a fairly long history in the astrophysical
literature.

Gravitational instability in infinitesimally thin discs has been
widely studied with fundamental contributions dating back to more than
fifty years ago \citep[e.g.][and references
  therein]{Lin64,Too64,Hun72}.  However, the thin-disc approximation
is not justified in many cases. In protostellar discs, the
  vertical extent of the gas can be substantial
  \citep[e.g.][]{Law22}.  Also gaseous discs in present-day galaxies
can have non-negligible thickness \citep{Yim14}, and there are
indications that disc thickness increases with redshift
\citep[][]{For06}, though dynamically cold disc are found also in
high-redshift galaxies \citep{Riz21}.  Thick gaseous discs are
  observed in present-day dwarf galaxies \citep[e.g.][]{Roy10} and
  expected in dwarf protogalaxies \citep{Nip15}. Moreover, the
  observational galactic volumetric star formation laws
  \citep[][]{Bac19b} strongly suggest that the 3D structure of discs
  has an important role in the process of conversion of gas into
  stars.

Several authors have tackled the problem of the gravitational
instability of non-razor-thin discs, essentially obtaining
modifications of the thin-disc stability criteria that account for
finite thickness
\citep[][]{Too64,Rom92,Ber10,Wan10,Elm11,Gri12,Rom13,Beh15}.  3D
systems were studied only under rather specific assumptions:
\citet{Cha61} analyzed infinite homogeneous rotating systems, while
\citet{Saf60}, \citet{Gen75} and \citet{Ber82} considered homogeneous
rotating slabs of finite thickness. \citet[][]{Gol65a,Gol65b}
accounted in detail for the vertical stratification of the gas
distribution, assuming polytropic equation of state (and thus
polytropic distributions) with specific values of the polytropic index
(see also \citealt{Mei22}).

In this work we address the general problem of the local gravitational
stability of rotating stratified fluids. Considering axisymmetric
perturbations, we derive 3D dispersion relations and stability
criteria for baroclinic and barotropic configurations, as well as for
somewhat idealized models of vertically stratified discs.

\section{Preliminaries}

We perform a linear stability analysis of rotating
astrophysical gaseous systems taking into account the self-gravity of
the perturbations. Here we introduce the equations on
which such analysis is based, and we define the general properties of
the unperturbed fluid and of the disturbances.

\subsection{Fundamental equations}

For our purposes the relevant set of equations consists in the
adiabatic inviscid fluid equations combined with the Poisson equation.
As it is natural when dealing with rotating system, we work in
cylindrical coordinates $(R,\phi,z)$. For simplicity we consider only
axisymmetric unperturbed configurations and perturbations, so all
derivatives with respect to $\phi$ are null. Under these assumptions
the fundamental system of equations reads
\begin{equation}\begin{split}
& \frac{\de \rho}{\de t} + \frac{1}{R}\frac{\de (R\rho \uR)}{\de R} + \frac{\de (\rho \uz)}{\de z} = 0 , \\
& \frac{\de \uR}{\de t} +\uR\frac{\de \uR}{\de R}+\uz\frac{\de \uR}{\de z}  - \frac{\uphi^2}{R} = -\frac{1}{\rho}\frac{\de \press}{\de R} - \frac{\de \Phi}{\de R}- \frac{\de \Phiext}{\de R},\\
& \frac{\de \uphi}{\de t} +\uR\frac{\de \uphi}{\de R}+\uz\frac{\de \uphi}{\de z} + \frac{\uR \uphi}{R}=0,\\
& \frac{\de \uz}{\de t} +\uR\frac{\de \uz}{\de R}+\uz\frac{\de \uz}{\de z} = -\frac{1}{\rho}\frac{\de \press}{\de z} - \frac{\de \Phi}{\de z}- \frac{\de \Phiext}{\de z},\\
    & \frac{\press}{\gamma-1}\left(\frac{\de}{\de t} + \uv\cdot\nabla\right) \ln (\press\rho^{-\gamma}) =0,\\
&\frac{1}{R}\frac{\de}{\de R}\left(R\frac{\de \Phi}{\de R}\right)+\frac{\de^2 \Phi}{\de z^2}=4\pi G\rho,
  \end{split}
\label{eq:axisymmeq}
\end{equation}
where, $\rho$, $\uv=(\uR,\uphi,\uz)$, $\press$, and $\Phi$ are,
respectively, the gas density, velocity, pressure and gravitational
potential, $\gamma$ is the adiabatic index, and $\Phiext$ is an
external {\em fixed} gravitational potential.

\subsection{Properties of the unperturbed system and of the perturbations}
\label{sec:prop_pert}

Let us consider a generic quantity $q=q(R,z,t)$ describing a property
of the fluid (such as $\rho$, $\press$, $\Phi$ or any component of
$\uv$): $q$ can be written as $q=\qunp+\delta q$, where the (time
independent) quantity $\qunp$ describes the stationary unperturbed
fluid and the (time dependent) quantity $\delta q$ describes the
Eulerian perturbation. From now on, without risk of ambiguity, we will
indicate any unperturbed quantity $\qunp$ simply as $q$.

We assume that the unperturbed system is a stationary rotating
($\uphi\neq0$) solution of \Eqs(\ref{eq:axisymmeq}) with no
meridional motions ($\uR=\uz=0$).  Limiting ourselves to a linear
stability analysis, we consider small ($|\delta q/q|\ll 1$) plane-wave
perturbations with spatial and temporal dependence $\delta q
\propto\exp[\im(\kR R+\kz z-\omega t)]$, where $\omega$ is the
frequency, and $\kR$ and $\kz$ are the radial and vertical components
of the wavevector $\kv$.

\section{Linear perturbation analysis and dispersion relations}
\label{sec:pert}

Here we present the linear analysis of the system~(\ref{eq:axisymmeq})
for a rotating stratified fluid perturbed with disturbances with
properties described in Section~\ref{sec:prop_pert}.  We derive the
dispersion relations for general baroclinic and barotropic
distributions, as well as for vertically stratified discs with
negligible radial density and pressure gradients.

\subsection{Baroclinic distributions}
\label{sec:pert_barocl}

When the unperturbed distribution is baroclinic, surfaces of constant
density and pressure do not coincide and $\Omega=\Omega(R,z)$, where
$\Omega$ is the angular velocity defined by $\uphi=\Omega R$.
Perturbing and linearizing \Eqs(\ref{eq:axisymmeq}), under the
assumption of a baroclinic unperturbed distribution, we get
\begin{equation}
  \begin{split}
    & - \im \omega \deltarho+\im\left(\kR-\frac{\im}{R}-\im\frac{\rhoR}{\rho}\right)\rho\delta\uR+\im\left(\kz-\im\frac{\rhoz}{\rho}\right)\rho\delta\uz=0,\\
    &-\im\omega\delta\uR-2\Omega\delta\uphi=-\im\frac{\kR}{\rho}\delta\press+\frac{\pressR}{\rho^2}\deltarho-\im\kR\delta\Phi,\\
  &-\im\omega\delta\uphi+\frac{\de (\Omega R)}{\de R}\delta\uR+R\frac{\de \Omega }{\de z}\delta\uz+\Omega\delta\uR=0,\\
    &-\im\omega\delta\uz=-\im\frac{\kz}{\rho}\deltapress+\frac{\pressz}{\rho^2}\deltarho-\im\kz\delta\Phi,\\
    &-\im\omega\frac{\deltapress}{\press}+\im\gamma\omega\frac{\delta\rho}{\rho}+\sigmaR\delta\uR+\sigmaz\delta\uz=0,\\
    &-\left(k^2-\im\frac{\kR}{R}\right)\delta\Phi=4\pi G\deltarho,\\
  \end{split}
  \label{eq:bcsys}
\end{equation}
where $k=\sqrt{\kR^2+\kz^2}$, $\rhoR\equiv\de \rho/\d R$,
$\rhoz\equiv\de \rho/\d z$, $\pressR\equiv\de \press/\de R$,
$\pressz\equiv\de \press/\de z$, $\sigmaR\equiv\de \sigma/\de R$,
$\sigmaz\equiv\de \sigma/\de z$ and
$\sigma\equiv\ln\left(\press\rho^{-\gamma}\right)$ is the normalized
specific entropy.  In deriving \Eqs(\ref{eq:bcsys}) we did not make
any assumption on the wavelength of the disturbance: assuming now that
$k$ is large compared to $1/R$, the system~(\ref{eq:bcsys}) leads to
the dispersion relation
\begin{equation}
\omega^4
+\left(4\pi G\rho-\kappa^2-\nu^2-\cs^2k^2\right)\omega^2
+\N^2 \cs^2k^2
+\cs^2\kz\left(\kz\kappa^2-\kR R\frac{\de \Omega^2}{\de z}\right)
-4\pi G \rho\frac{\kz}{k^2}\left(\kz\kappa^2-\kR R\frac{\de \Omega^2}{\de z}\right)
+\kappa^2\nu^2=0,
\label{eq:bc}
\end{equation}
where $\cs^2=\gamma\press/\rho$ is the adiabatic sound speed squared, $\kappa$
is the epicycle frequency, defined by
\begin{equation}  
\kappa^2\equiv 4\Omega^2+\frac{\d \Omega^2}{\d \ln R},
\end{equation}
\begin{equation}
  \N^2\equiv -\frac{1}{\gamma\rho}
  \left[\frac{\kz^2}{k^2}\sigmaR\pressR+\frac{\kR^2}{k^2}\sigmaz\pressz-\frac{\kR\kz}{k^2}\left(\sigmaR\pressz+\sigmaz\pressR\right)\right] 
\label{eq:bvfreq}
\end{equation}
is a generalized buoyancy (or Brunt-V\"ais\"al\"a) frequency squared
\citep[see][]{Bal95}, and we have introduced the frequency $\nu$,
defined by
\begin{equation}  
\nu^2\equiv \frac{\rhoz\pressz}{\rho^2}=\frac{\cs^2}{\gamma}\frac{\rhoz}{\rho}\frac{\pressz}{\press},
\label{eq:nusquared}
\end{equation}
which is related to vertical pressure and density gradients.

\subsection{Barotropic distributions}
\label{sec:per_barotr}

When the unperturbed distribution is barotropic, the isobaric and
isopycnic surfaces coincide, and $\Omega=\Omega(R)$
\citep[e.g.][]{Tas78}.  The dispersion relation for the barotropic
case, obtained from \Eq(\ref{eq:bc}) substituting $\de \Omega^2/\de
z=0$, is
\begin{equation}
  \omega^4
  +\left(4\pi G\rho-\kappa^2-\nu^2-\cs^2k^2\right)\omega^2
+\N^2\cs^2k^2 
  +\kappa^2\cs^2\kz^2
  -4\pi G \rho \kappa^2\frac{\kz^2}{k^2}
+\kappa^2\nu^2=0,
\label{eq:bt}
\end{equation}
where, given that $\press=\press(\rho)$, $\N^2$ can be written as
\begin{equation}
\N^2= -\frac{1}{\gamma\rho}\frac{\d \sigma}{\d\rho}\frac{\d \press}{\d \rho}\left(\frac{\kR}{k}\rhoz-\frac{\kz}{k}\rhoR\right)^2.
\label{eq:bvfreqbt}
\end{equation}

\subsection{Vertically stratified discs}
\label{sec:pert_thick}

A gaseous disc with finite thickness can be approximately described
over most of its radial extent by a stationary rotating fluid with
negligible radial gradients of pressure and density compared to the
corresponding vertical gradients. If we further assume that
$\Omega=\Omega(R)$, the dispersion relation describing the evolution
of axisymmetric perturbations in such a disc model can be obtained
from \Eqs(\ref{eq:bt}) and (\ref{eq:bvfreqbt}), simply by
imposing\footnote{For the stationary hydrodynamic equations to be
  satisfied with $\Omega=\Omega(R)$ and $\rhoR=0$, the gravitational
  potential must be separable in cylindrical coordinates. Though in
  general this is not the case globally, it can be locally a
  reasonable approximation for our idealized model.}  $\rhoR=0$. The
resulting dispersion relation is
\begin{equation}
  \omega^4
  +\left(4\pi G\rho-\kappa^2-\nu^2-\cs^2k^2\right)\omega^2
+\Nz^2\cs^2\kR^2 
  +\kappa^2\cs^2\kz^2
  -4\pi G \rho \kappa^2\frac{\kz^2}{k^2}
+\kappa^2\nu^2=0,
\label{eq:thick}
\end{equation}
where $\Nz^2\equiv-\sigmaz\pressz/(\gamma\rho)$ is the vertical
Brunt-V\"ais\"al\"a frequency squared.

\section{Stability criteria}
\label{sec:crit}

Here we derive the stability criteria obtained analyzing the
dispersion relations of Section~\ref{sec:pert}, starting from the
simplest case (vertically stratified discs) and then moving to more
general barotropic and baroclinic distributions. The dispersion
relations of Section~\ref{sec:pert} were derived without any
assumption on the sign of $\kappa^2$, $\Nz^2$ and $\nu^2$.  However,
given that we are interested in the gravitational instabilities, in
the following we perform the stability analysis assuming $\kappa^2>0$
and $\Nz^2>0$, to exclude rotational and convective instabilities, at
least when $\Omega=\Omega(R)$ \citep[e.g.][]{Tas78}.  It is useful to
note that
\begin{equation}
  \Nz^2=
  -\frac{\pressz}{\gamma\rho}\left(\frac{\pressz}{\press}-\gamma\frac{\rhoz}{\rho}\right)=
  \nu^2-\frac{(\pressz)^2}{\gamma\rho\press}<\nu^2,
\label{eq:nznu}
\end{equation}
so our assumption $\Nz^2>0$ implies $\nu^2>0$. 

The dispersion relations of Section~\ref{sec:pert} were derived using
as only assumption on the perturbation wavenumber that $k$ is larger
than $1/R$.  Further restrictions on $k$ derive from the requirement
that the size of the disturbance is smaller than the characteristic
length scales of the unperturbed system. Thus, based on the arguments
reported in Appendix~\ref{sec:app_local}, the following stability
analysis (with the only exception of Section~\ref{sec:more_general_modes}) will be restricted to modes with
\begin{equation}
k^2>\kJ^2+\frac{\nu^2}{\cs^2},
\label{eq:k_restriction}
\end{equation}
where $\kJ=\sqrt{4\pi G\rho}/\cs$ is the Jeans wavenumber. In
Section~\ref{sec:more_general_modes}, where the analysis is limited to
radial modes in vertically stratified discs, we consider also
longer-wavelength modes that do not satisfy
inequality~(\ref{eq:k_restriction}).

\subsection{Criteria for vertically stratified discs}
\label{sec:crit_thick}

Using the notation introduced at the beginning of
Appendix~\ref{sec:app_disp_rel}, the dispersion
relation~(\ref{eq:thick}) can be written in the form of
\Eq(\ref{eq:disp_abcdzeta}).  Analyzing this dispersion relation,
  in Section~\ref{sec:disp_abcdzeta} we show that for vertically
  stratified discs a sufficient condition for stability is
  inequality~(\ref{eq:suff_stab_abcdzeta}), i.e.\
\begin{equation}
4\pi G\rho \Nz^2<(\nu^2-\Nz^2)(\Nz^2-\kappa^2)\qquad\mbox{(sufficient for stability)}.
\label{eq:suff_stab_thick}  
\end{equation}
We recall that this criterion refers only to stability against
short-wavelength perturbations (i.e.\ modes satisfying the condition
\ref{eq:k_restriction}), so stability against longer-wavelength modes
is not guaranteed.  In the following we analyze the behavior of
specific families of modes, which could allow us to obtain sufficient
criteria for instability.

\subsubsection{Modes with $\kR=0$}
\label{sec:vertical_modes}

For vertical modes the dispersion relation, obtained substituting
$\kR=0$ in \Eq(\ref{eq:thick}), is
\begin{equation}
  \omega^4
  +\left(4\pi G\rho-\kappa^2-\nu^2-\cs^2\kz^2\right)\omega^2
  +\kappa^2\cs^2\kz^2
  -4\pi G \rho \kappa^2+\kappa^2\nu^2=0,  
\label{eq:thickvert}
\end{equation}
which is in the form of \Eq(\ref{eq:disp_abc}).  In
Section~\ref{sec:disp_abc} we show that all vertical modes satisfying
condition~(\ref{eq:k_restriction}) are stable.


\subsubsection{Modes with $\kz=0$}
\label{sec:more_general_modes}

Given that our disc model has no density or pressure radial gradients,
when studying purely radial ($\kz=0$) modes we can relax the
assumption~(\ref{eq:k_restriction}), so we consider here also smaller
$|\kR|$ modes, requiring only\footnote{We must also require that
  $|\kR|$ is larger than $(\d \kappa^2/\d R)/\kappa^2$, which however
  is typically of the order of $1/R$.} that $|\kR|$ is larger than
$1/R$.  However, as pointed out by \citet{Saf60} and \citet{Gol65a},
when considering radial modes with $|\kR|$ smaller than $\sim 1/\hz$,
where $\hz$ is the disc thickness, care must be taken when perturbing
the Poisson equation, to avoid the unphysical divergence for small
$|\kR|$ that one would obtain from the last equation of
system~(\ref{eq:bcsys}) when $\kz=0$. Following \citet{Gol65a}, here
we consider a perturbed Poisson equation in the form
\begin{equation}
 -\left(\kR^2+\hz^{-2}\right)\delta\Phi=4\pi G\deltarho,
\label{eq:poisson_hz}
\end{equation}  
which approximately accounts for the finite vertical extent of the
disc \citep[see also][for similar approaches in 2D
  models]{Too64,Shu68,Van70,Yue82}.  Combining this equation with the
first five equations of system~(\ref{eq:bcsys}), assuming $\kz=0$ and
$\rhoR=\pressR=\sigmaR=0$, for wavenumbers larger than $1/R$ we get
the dispersion relation\footnote{When $\rhoz=0$ (and thus $\Nz=0$ and
  $\nu=0$), this dispersion relation reduces to a quadratic dispersion
  relation, which is essentially that obtained by \citet{Saf60} for a
  homogeneous disc of finite thickness.}
\begin{equation}
  \omega^4
  +\left(4\pi G\rho\frac{\kR^2}{\kR^2+\hz^{-2}}-\kappa^2-\nu^2-\cs^2\kR^2\right)\omega^2
+\Nz^2\cs^2\kR^2 
+\kappa^2\nu^2=0,
\label{eq:thickrad}
\end{equation}
which is in the form of \Eq(\ref{eq:disp_abcde}). In
Section~\ref{sec:disp_abcde} we show that for this dispersion relation
a sufficient condition for instability is
inequality~(\ref{eq:crit_rad_mod}), which can be rewritten as
\begin{equation}
\Qthreed\equiv\frac{\sqrt{\kappa^2+\nu^2}+\cs\hz^{-1}}{\sqrt{4\pi G\rho}}<1\;\;\;\;\mbox{(sufficient for instability)}.
\label{eq:suff_instab_thick}  
\end{equation}
When this condition is satisfied the instability occurs for
intermediate values of $|\kR|$, i.e.\ those modes that satisfy
inequality~(\ref{eq:two_crit_s}), i.e.\
\begin{equation}
  \cs^4\kR^4-\left(4\pi G\rho-\kappa^2-\nu^2-\cs^2\hz^{-2}\right)\cs^2\kR^2+\cs^2\hz^{-2}(\kappa^2+\nu^2)<0,
\label{eq:two_crit_k}  
\end{equation}
consistent with the general finding that the short-wavelength
disturbances are stabilized by pressure and long-wavelength
disturbances by rotation\footnote{For an infinite homogeneous
  uniformly rotating medium ($\hz\to\infty$, $\nu=0$,
  $\kappa^2=4\Omega^2$), condition~(\ref{eq:two_crit_k}) reduces to
  $\cs^2\kR^2<4\pi G\rho-4\Omega^2$, which is the instability criterion
  found by \citet{Cha61}.}  \citep[e.g.][]{Too64,Gol65a}.

To gauge the parameter $\hz$ appearing in
\Eqs(\ref{eq:poisson_hz}-\ref{eq:two_crit_k}), in
Appendix~\ref{sec:app_hz} we compare the
criterion~(\ref{eq:suff_instab_thick}) with those obtained for two
specific models by \citet{Gol65a}. This comparison suggests to adopt
$\hsixty\lesssim \hz\lesssim \heighty$, where $\hXperc$ is the height
of a strip centred on the midplane containing $X\%$ of the mass per
unit surface. $\hz\approx\hseventy$ can be taken as reference fiducial
value.


\subsection{Criteria for barotropic distributions}
\label{sec:crit_barot}

We consider here the dispersion relation~(\ref{eq:bt}) obtained for
barotropic distributions.  We did attempt to analyze this dispersion
relation with an approach similar to that of
Section~\ref{sec:disp_abcdzeta}, but we did not find simple general
stability criteria independent of the wavevector.  However, as in the
case of vertically stratified discs
(Section~\ref{sec:more_general_modes}), a sufficient criterion for
instability can be obtained by considering purely radial
perturbations. When $\kz=0$ the dispersion relation~(\ref{eq:bt}) for
barotropic distributions becomes
\begin{equation}
  \omega^4
  +\left(4\pi G\rho-\kappa^2-\nu^2-\cs^2\kR^2\right)\omega^2
+\Nz^2\cs^2\kR^2 
+\kappa^2\nu^2=0,
\label{eq:bt_rad}
\end{equation}
which is in the form of \Eq(\ref{eq:disp_abcd}).  In
Section~\ref{sec:disp_abcd} we show that a sufficient condition for
instability is inequality~(\ref{eq:suff_instab_abcd}), which can be
rewritten as
\begin{equation}
4\pi G\rho \Nz^2> \nu^2(\nu^2-\Nz^2)+(\kappa^2/2)^2\qquad\mbox{(sufficient for instability)}.
\label{eq:suff_instab_bt_rad}
\end{equation}
We note that \Eq(\ref{eq:nznu}) implies that the r.h.s.\ of this
inequality is always positive, so stratification, as well as rotation,
can contribute to counteract the instability.



\subsection{Criteria for baroclinic distributions}
\label{sec:crit_baroc}

The dispersion relation found for baroclinic distributions
(\Eq\ref{eq:bc}) differs from the corresponding dispersion relation
for barotropic distributions (\Eq\ref{eq:bt}) only for the presence of
terms $\propto \kz\kR \de \Omega^2/\de z$. Thus, the behavior of
radial ($\kz=0$) modes in baroclinic distributions is determined by
the dispersion relation~(\ref{eq:bt_rad}).  It follows that the
sufficient criterion for instability~(\ref{eq:suff_instab_bt_rad})
applies also to systems with baroclinic distributions.

\section{A case study: discs in vertical isothermal equilibrium}
\label{sec:case_study}

As a case study, we consider here a simple model of a disc with the
properties described in Section~\ref{sec:pert_thick}, without external
potential, assuming that the vertical density distribution is given by
the self-gravitating isothermal slab \citep{Spi42}
\begin{equation}
\rho(z)=\rhozero\sech^2\ztilde,
\label{eq:rho_slab}
\end{equation}  
where $\rhozero=\rho(0)$ is the density in the midplane,
$\ztilde\equiv z/b$ and $b=\csiso/\sqrt{2\pi G\rhozero}$, where
$\csiso\equiv\cs\gamma^{-1/2}$ is the position-independent isothermal
sound speed (in this section we assume $\gamma=5/3$).  For this model
we have
\begin{equation}
\nu^2=8\pi G \rhozero\tanh^2\ztilde
\label{eq:nu_slab}
\end{equation}
and
\begin{equation}
\Nz^2=\frac{2}{5}\nu^2.  
\label{eq:nz_slab}
\end{equation}

\subsection{Sufficient criterion for instability}
\label{sec:case_study_inst}

\begin{figure}
  \centerline{\includegraphics[width=0.5\textwidth]{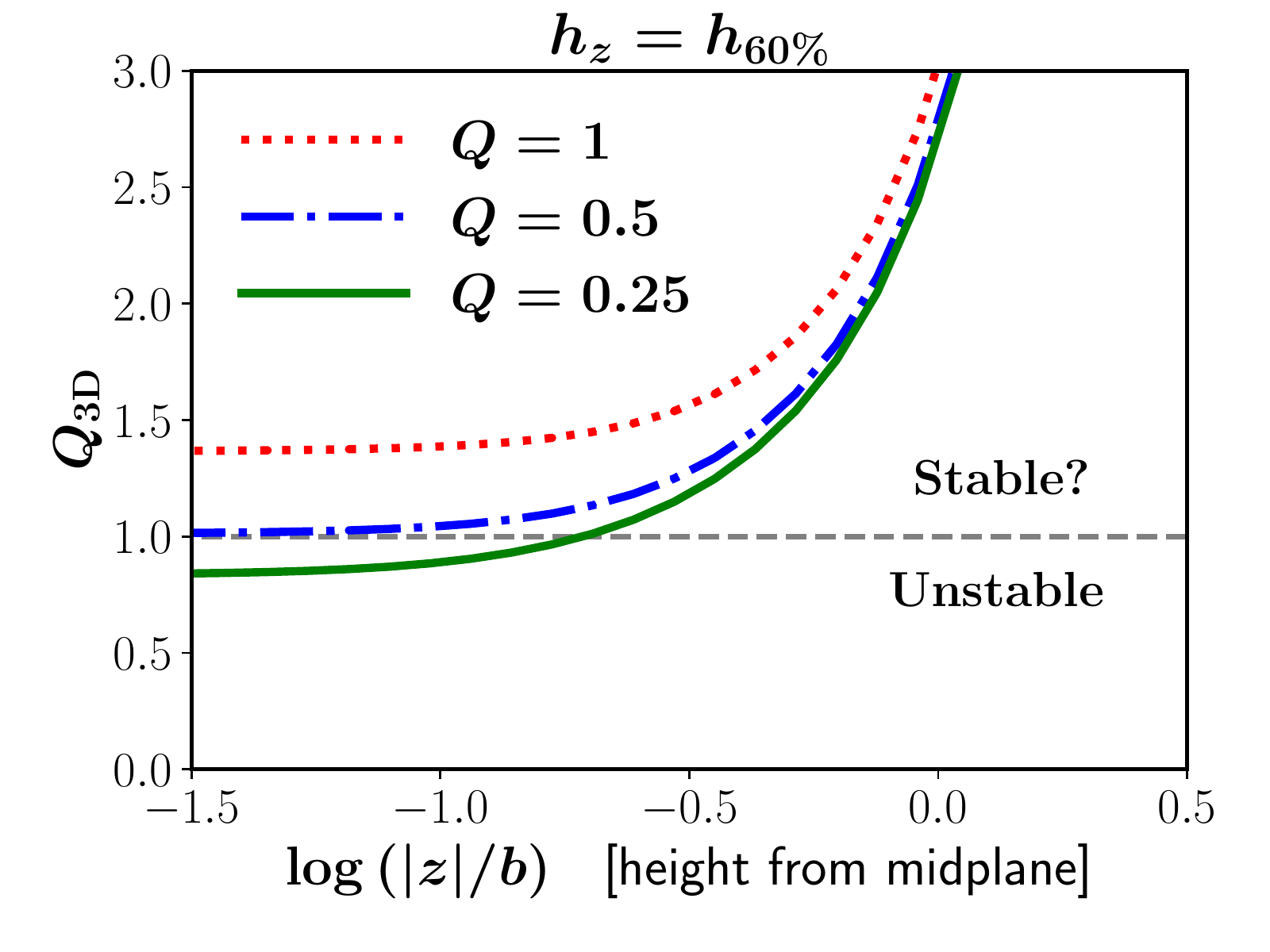}\includegraphics[width=0.5\textwidth]{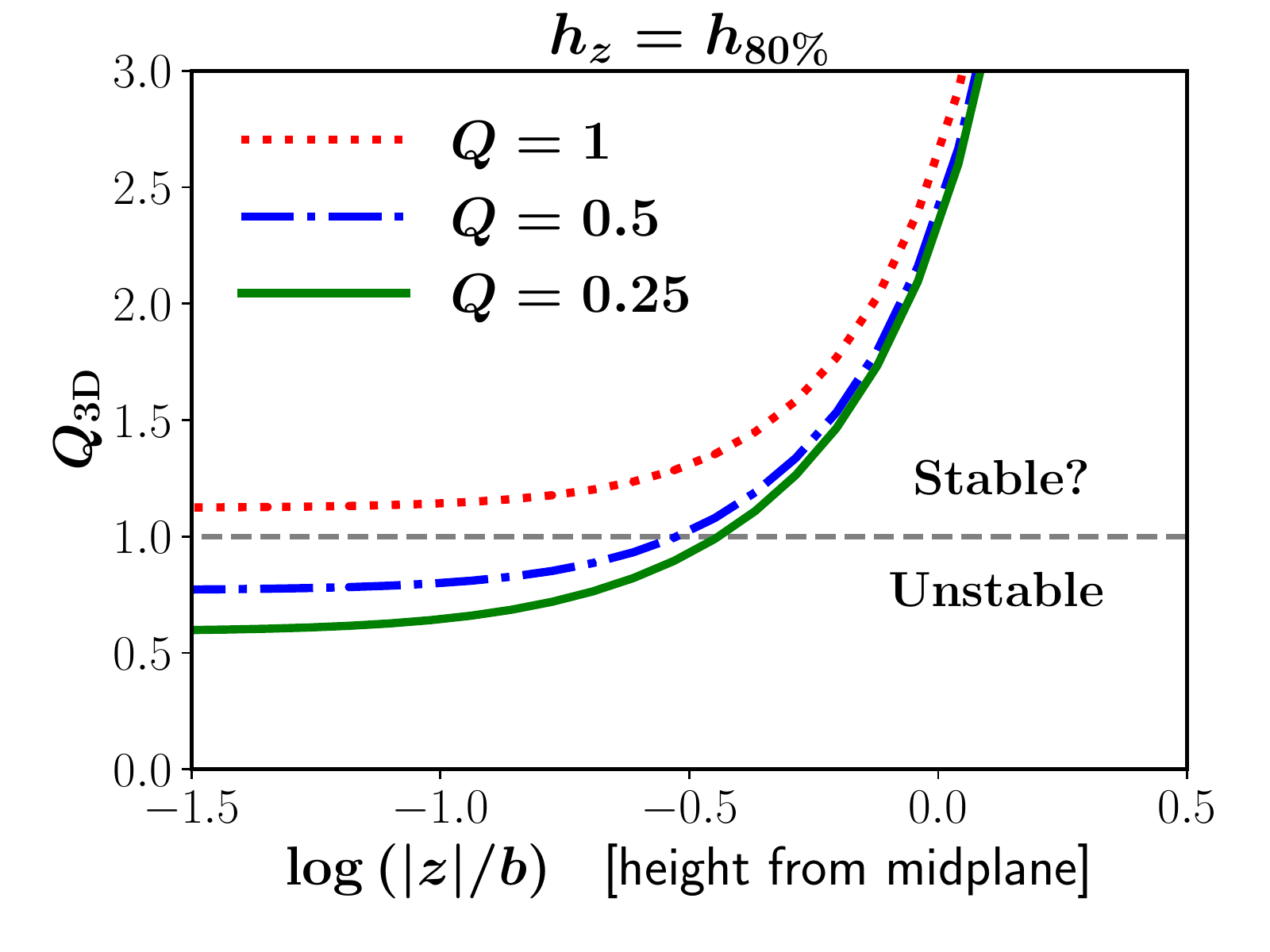}}
\caption{The 3D instability parameter $\Qthreed$ as a function of
    height from the midplane $|z|$ for a rotating, self-gravitating
    stratified disc in vertical isothermal equilibrium
    (\Eq\ref{eq:rho_slab}) at a given radius, for different values of
    Toomre's 2D instability parameter $Q$. $\Qthreed$ is calculated
    for $\gamma=5/3$ and either $\hz=\hsixty$ (thickness containing
    60\% of the mass per unit surface; left panel) or $\hz=\heighty$
    (thickness containing 80\% of the mass per unit surface; right
    panel). Our sufficient 3D instability criterion predicts
    instability where $\Qthreed<1$, but does not guarantee stability
    where $\Qthreed>1$. \label{fig:q3d}}
\end{figure}

Using \Eqs(\ref{eq:rho_slab}-\ref{eq:nz_slab}) and $\rhozero=\pi
G\Sigma^2/(2\csiso^2)$, where
\begin{equation}
\Sigma=\int_{-\infty}^\infty\rho(z)\d z 
\label{eq:surf_den}
\end{equation}
is the surface density, for the isothermal disc the sufficient
condition for instability (\ref{eq:suff_instab_thick}) becomes
\begin{equation}
  \Qthreed=\sqrt{\frac{Q^2}{2}\cosh^2\ztilde+2\sinh^2\ztilde}
  +\sqrt{\frac{5}{6}}\frac{b}{\hz}\cosh\ztilde<1\qquad\mbox{(sufficient
    for instability)},
\label{eq:q3d_slab}
\end{equation}
where
\begin{equation}
  Q\equiv\frac{\kappa\csiso}{\pi G\Sigma}
\end{equation}
is the classical 2D \citet{Too64} instability parameter at a given
radius. Fig.~\ref{fig:q3d} shows $\Qthreed$ as a function of $z$ for
representative values of $Q$ when $\hz=\hsixty\simeq 1.4b$ (left
panel) and $\hz=\heighty\simeq 2.2b$ (right panel), which should
bracket realistic values of $\hz$ (see
Section~\ref{sec:more_general_modes} and
Appendix~\ref{sec:app_hz}). $\Qthreed$ is an increasing function of
$|z|$, so, at given radius, the disc is more prone to gravitational
instability near the midplane.  For both choices of $\hz$ the $Q=1$
model is stable and the $Q=0.25$ model is unstable; the $Q=0.5$ model
is marginally stable for $\hz=\hsixty$ and unstable for
$\hz=\heighty$.  The overall condition to have instability at any
height at a given radius in the considered stratified disc is
$\Qthreedmin=\Qthreed(0)<1$, i.e.\
\begin{equation}
Q<\Qcrit=\sqrt{2}-\sqrt{\frac{5}{3}}\frac{b}{\hz} \qquad\mbox{(sufficient for instability)},
\label{eq:qthreedmin}
\end{equation}
which gives $\Qcrit\simeq0.48$ for $\hz=\hsixty$ and $\Qcrit\simeq
0.83$ for $\hz=\heighty$, broadly consistent with Toomre's 2D
criterion $Q<1$, given the known stabilizing effect of finite
thickness (see Section~\ref{sec:corrected2d}).

When the conditions for instability are met, there is a range of
unstable radial wavenumbers (satisfying inequality
\ref{eq:two_crit_k}) centred at $|\kR|=\kRinst$ (see
Appendix~\ref{sec:disp_abcde}). For the discs here considered
$\kRinst$ is largest in the midplane, where it can be written as
  \begin{equation}
\kRinst^2\hz^2=\frac{3}{5}\frac{\hz^2}{b^2}\left(1-\frac{Q^2}{2}\right)-\frac{1}{2},
  \end{equation}  
which gives $0.7\lesssim \kRinst\hz\lesssim 1.5$ for $\hsixty\lesssim
\hz \lesssim \heighty$ and $0\lesssim Q\lesssim \Qcrit$.  Thus, the
typical unstable modes ($|\kR|\hz\approx 1$) have radial wavelength
$\approx 2\pi\hz$, consistent with estimates obtained in finite
thickness-corrected 2D models (\citealt{Kim02,Rom14,Beh15}; see
Section~\ref{sec:corrected2d}).  We note that, provided that $\hz$ is
smaller than $R$, $\kRinst$ is larger than $1/R$, consistent with our
assumptions.


\subsection{Sufficient criterion for stability}
\label{sec:case_study_stab}

\begin{figure}
  \centerline{\includegraphics[width=0.5\textwidth]{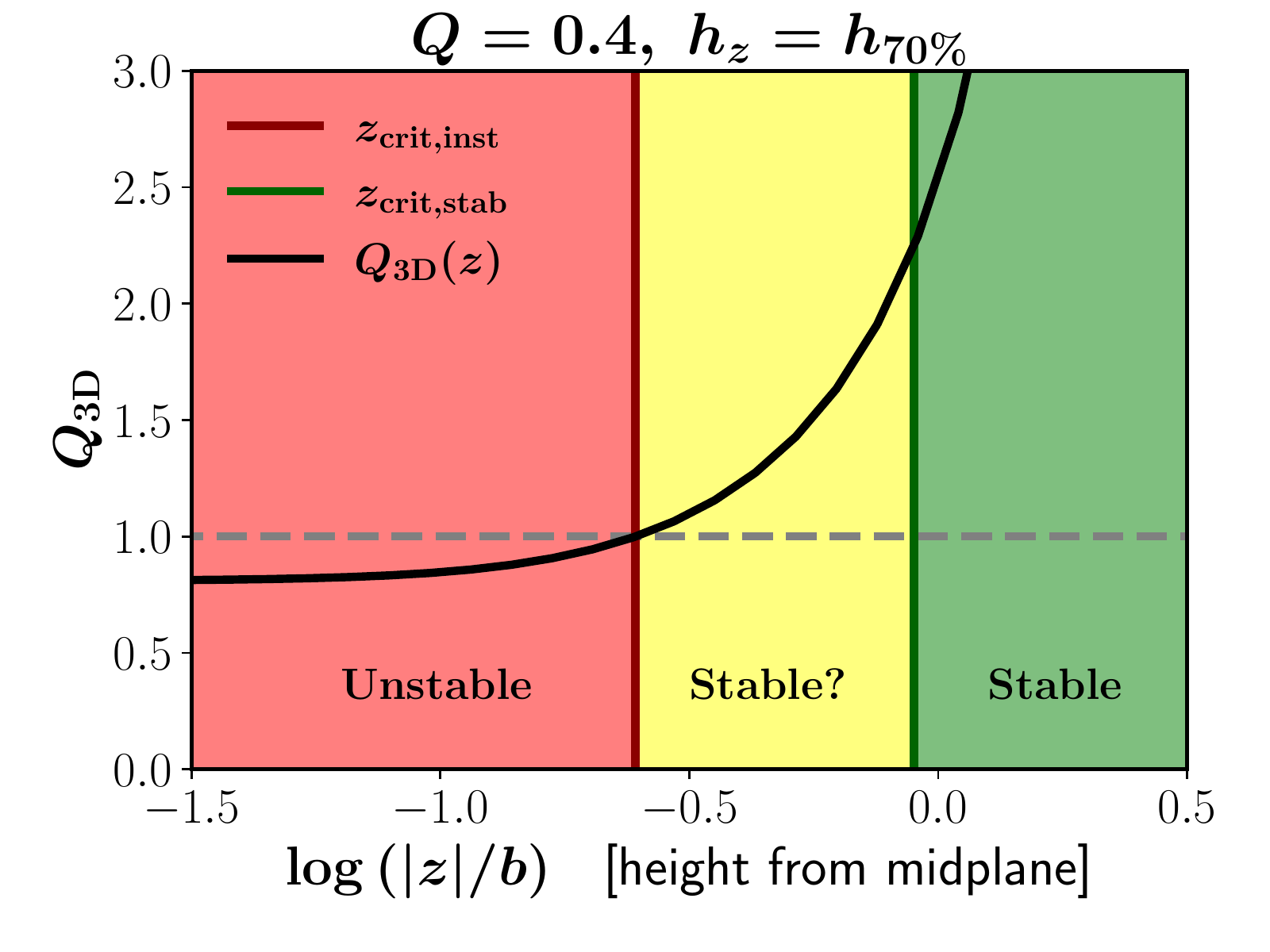}}
  \caption{Instability and stability regions as a function of height
    from the midplane $|z|$ at a given radius for a rotating,
    self-gravitating stratified disc in vertical isothermal
    equilibrium (\Eq\ref{eq:rho_slab}) with 2D Toomre's parameter
    $Q=0.4$ for $\gamma=5/3$. The instability region
    ($|z|<\zcritinst$; red) is determined by the criterion
    $\Qthreed<1$ (\Eq\ref{eq:q3d_slab}) with $\hz=\hseventy$. The
    stability region ($|z|>\zcritstab$; green) is determined by the
    condition~(\ref{eq:stab_slab}). In the yellow region stability is
    not guaranteed by our criteria.
    \label{fig:stab_instab}}
\end{figure}

Combining \Eq(\ref{eq:suff_stab_thick}) with \Eqs(\ref{eq:nu_slab})
and (\ref{eq:nz_slab}), we get that for the isothermal stratified disc
a sufficient condition for stability is
\begin{equation}
\frac{4}{3\cosh^2\ztilde}\left(\frac{6}{5}\sinh^2\ztilde-1\right)>Q^2\qquad\mbox{(sufficient
  for stability)}.
\label{eq:stab_slab}
\end{equation}
The l.h.s.\ of this inequality is an increasing function of $|z|$.
When $Q\lesssim 1.27$ this sufficient criterion guarantees stability
at $|z|>\zcritstab$, where
$\zcritstab>b\asinh{\sqrt{5/6}}\simeq 0.82b$ is an increasing
function of $Q$.

Fig.\ \ref{fig:stab_instab} shows, for a representative isothermal
disc with $Q=0.4$ at a given radius, stability and instability
regions, as a function of height from midplane, obtained combining our
sufficient criteria for stability (\Eq\ref{eq:stab_slab}) and
instability (\Eq\ref{eq:q3d_slab}), taking for the latter as fiducial
value $\hz=\hseventy\simeq 1.7 b$, such that $\Qcrit\simeq 0.67$
(\Eq\ref{eq:qthreedmin}). The disc is unstable close to the midplane
(at $|z|<\zcritinst\simeq 0.25b$) and stable at $|z|>\zcritstab\simeq
0.9b$, while stability is not guaranteed at intermediate heights.

\subsection{Comparison with finite-thickness corrected 2D models}
\label{sec:corrected2d}

Here we compare our results on the isothermal disc with those obtained
with modifications of the thin-disc stability criteria that account
for finite thickness, which, as mentioned in Section~\ref{sec:intro},
is a complementary approach to study the local gravitational
instability in realistically thick discs.  These modified 2D criteria
are based on 2D models in which the self-gravity of the perturbation
is corrected with a reduction factor $\F$, depending both on the
radial wavenumber of the perturbation and on the disc scale height
$h$. Different authors have adopted different functional forms of
$\F$, but, for given $\F$, by computing the wavenumber of the most
unstable mode, it is always possible to express the condition for
instability as $Q<\Qcrit$, where $\Qcrit$ depends on the unperturbed
vertical density distribution. Focusing on the self-gravitating
isothermal disc, we can thus compare the values of $\Qcrit$ that we
find using our 3D criterion ($\Qcrit\simeq 0.5,\,0.7$ and $0.8$ for
$\hz=\hsixty,\,\hseventy$ and $\heighty$, respectively; see
Sections~\ref{sec:case_study_inst} and \ref{sec:case_study_stab}),
with those obtained in the literature using modified 2D criteria:
$\Qcrit\simeq 0.65$ \citep{Kim02}, $0.6\lesssim \Qcrit \lesssim 0.65$
(\citealt{Ber10}, considering two different functional forms of $\F$;
see also \citealt{Ber14}), $\Qcrit\simeq 0.69$ \citep{Wan10},
$\Qcrit\simeq 0.67$ (\citealt{Rom13}, based on the calculations
presented in \citealt{Rom92}) and $\Qcrit\simeq 0.70$ \citep{Beh15}.
These values of $\Qcrit$ are consistent with those found with our 3D
criterion, with a remarkably good agreement when $\hz\approx
\hseventy$.

\section{Conclusions}

In this letter we have derived dispersion relations for axisymmetric
perturbations, which can be used to study the local gravitational
stability in stratified rotating axisymmetric gaseous systems with
general baroclinic (\Eq\ref{eq:bc}) and barotropic (\Eq\ref{eq:bt})
distributions, as well as in vertically stratified discs
(\Eqs\ref{eq:thick}, \ref{eq:thickvert} and \ref{eq:thickrad}).  We
have obtained 3D sufficient stability (\Eq\ref{eq:suff_stab_thick})
and instability (\Eqs\ref{eq:suff_instab_thick} and
\ref{eq:suff_instab_bt_rad}) criteria, which generalize previous
results and can be used to determine whether and where a rotating
system of given 3D structure is prone to fragmentation and clump
formation.

In the case of vertically stratified discs, we have expressed the
sufficient instability criterion as $\Qthreed<1$
(\Eq\ref{eq:suff_instab_thick}), where the dimensionless parameter
$\Qthreed=\left(\sqrt{\kappa^2+\nu^2}+\cs\hz^{-1}\right)/{\sqrt{4\pi
    G\rho}}$ can be seen as a 3D version of Toomre's 2D $Q$ parameter,
in which the combined stabilizing effects of rotation ($\kappa^2$) and
stratification ($\nu^2$) are apparent.  A shortcoming of this 3D
criterion is that the disc thickness parameter $\hz$ is not exactly
defined.  However, the comparison with previous 2D and 3D models in
the literature (Section~\ref{sec:case_study} and
Appendix~\ref{sec:app_hz}) suggests to use as fiducial value
$\hz\approx \hseventy$, where $\hseventy$ is the height of a strip
centred on the midplane containing 70\% of the mass per unit surface.
Independent of the specific assumed definition of $\hz$, applying our
criteria to discs with isothermal vertical stratification, we have
shown quantitatively that the conditions for gravitational instability
are more easily met close to the midplane, while stability prevails
far from the midplane. In the midplane of unstable discs, the typical
perturbations that are expected to grow have radial extent of a few
$\hz$.

When the conditions for gravitational instability are satisfied, the
perturbations are expected to grow and enter the non-linear regime,
which cannot be studied using the linearized equations considered in
this work. Though numerical simulations would be necessary to describe
quantitatively the non-linear growth of axisymmetric disturbances,
qualitatively we expect that the outcome of the instability would be
the formation of thick ring-like structures in the equatorial plane of
the rotating gaseous systems, which might then fragment into spiral
arms, filaments and clumps \citep[see][]{Wan10,Beh15}. These clumps
are likely to be the sites of star formation in galactic discs and
possibly of planet formation in protostellar discs.  Collapsed
overdense rings are not expected to form out of the midplane, not only
because there the instability conditions are harder to meet (see
Figs~\ref{fig:q3d} and \ref{fig:stab_instab}), but also because in the
vertical direction the gravitational instability is essentially
Jeans-like with Jeans length is of the order of the vertical
scale-height (see Section~\ref{sec:vertical_modes} and
Appendix~\ref{sec:app_local}), so there is no room to form vertically
distinct rings.  The 3D structure of filaments formed in the midplane
of gravitationally unstable plane-parallel stratified non-rotating
systems has been studied with hydrodynamic simulations by \citet[][see
  their figures 6 and 7]{van14}. Mutatis mutandis, the results of
\citet{van14} suggest that, in an unstable rotating stratified disc,
the collapsed overdense rings will likely have vertical density
distributions similar in shape to that of the unperturbed disc, but
with smaller scaleheight.




\section*{Acknowledgements}

I am grateful to the anonymous referee for useful suggestions that
helped improve the paper.

\section*{Data Availability}

The data underlying this article will be shared on reasonable request
to the corresponding author.

\bibliography{biblio_gri}

\begin{thebibliography}{}
\makeatletter
\relax
\def\mn@urlcharsother{\let\do\@makeother \do\$\do\&\do\#\do\^\do\_\do\%\do\~}
\def\mn@doi{\begingroup\mn@urlcharsother \@ifnextchar [ {\mn@doi@}
  {\mn@doi@[]}}
\def\mn@doi@[#1]#2{\def\@tempa{#1}\ifx\@tempa\@empty \href
  {http://dx.doi.org/#2} {doi:#2}\else \href {http://dx.doi.org/#2} {#1}\fi
  \endgroup}
\def\mn@eprint#1#2{\mn@eprint@#1:#2::\@nil}
\def\mn@eprint@arXiv#1{\href {http://arxiv.org/abs/#1} {{\tt arXiv:#1}}}
\def\mn@eprint@dblp#1{\href {http://dblp.uni-trier.de/rec/bibtex/#1.xml}
  {dblp:#1}}
\def\mn@eprint@#1:#2:#3:#4\@nil{\def\@tempa {#1}\def\@tempb {#2}\def\@tempc
  {#3}\ifx \@tempc \@empty \let \@tempc \@tempb \let \@tempb \@tempa \fi \ifx
  \@tempb \@empty \def\@tempb {arXiv}\fi \@ifundefined
  {mn@eprint@\@tempb}{\@tempb:\@tempc}{\expandafter \expandafter \csname
  mn@eprint@\@tempb\endcsname \expandafter{\@tempc}}}

\bibitem[\protect\citeauthoryear{{Bacchini}, {Fraternali}, {Pezzulli},
  {Marasco}, {Iorio}  \& {Nipoti}}{{Bacchini} et~al.}{2019}]{Bac19b}
{Bacchini} C.,  {Fraternali} F.,  {Pezzulli} G.,  {Marasco} A.,  {Iorio} G.,
  {Nipoti} C.,  2019, \mn@doi [\aap] {10.1051/0004-6361/201936559}, \href
  {https://ui.adsabs.harvard.edu/abs/2019A&A...632A.127B} {632, A127}

\bibitem[\protect\citeauthoryear{{Balbus}}{{Balbus}}{1995}]{Bal95}
{Balbus} S.~A.,  1995, \mn@doi [\apj] {10.1086/176397}, \href
  {https://ui.adsabs.harvard.edu/abs/1995ApJ...453..380B} {453, 380}

\bibitem[\protect\citeauthoryear{{Behrendt}, {Burkert}  \&
  {Schartmann}}{{Behrendt} et~al.}{2015}]{Beh15}
{Behrendt} M.,  {Burkert} A.,   {Schartmann} M.,  2015, \mn@doi [\mnras]
  {10.1093/mnras/stv027}, \href
  {https://ui.adsabs.harvard.edu/abs/2015MNRAS.448.1007B} {448, 1007}

\bibitem[\protect\citeauthoryear{{Bertin}}{{Bertin}}{2014}]{Ber14}
{Bertin} G.,  2014, {Dynamics of Galaxies}.
Cambridge University Press

\bibitem[\protect\citeauthoryear{{Bertin} \& {Amorisco}}{{Bertin} \&
  {Amorisco}}{2010}]{Ber10}
{Bertin} G.,  {Amorisco} N.~C.,  2010, \mn@doi [\aap]
  {10.1051/0004-6361/200913611}, \href
  {https://ui.adsabs.harvard.edu/abs/2010A&A...512A..17B} {512, A17}

\bibitem[\protect\citeauthoryear{{Bertin} \& {Casertano}}{{Bertin} \&
  {Casertano}}{1982}]{Ber82}
{Bertin} G.,  {Casertano} S.,  1982, \aap, \href
  {https://ui.adsabs.harvard.edu/abs/1982A&A...106..274B} {106, 274}

\bibitem[\protect\citeauthoryear{{Binney} \& {Tremaine}}{{Binney} \&
  {Tremaine}}{2008}]{Bin08}
{Binney} J.,  {Tremaine} S.,  2008, {Galactic Dynamics: Second Edition}.
Princeton University Press

\bibitem[\protect\citeauthoryear{{Chandrasekhar}}{{Chandrasekhar}}{1961}]{Cha61}
{Chandrasekhar} S.,  1961, {Hydrodynamic and hydromagnetic stability}.
Clarendon Press: Oxford University Press

\bibitem[\protect\citeauthoryear{{Cimatti}, {Fraternali}  \&
  {Nipoti}}{{Cimatti} et~al.}{2019}]{CFN19}
{Cimatti} A.,  {Fraternali} F.,   {Nipoti} C.,  2019, {Introduction to galaxy
  formation and evolution: from primordial gas to present-day galaxies}.
Cambridge University Press

\bibitem[\protect\citeauthoryear{{Elmegreen}}{{Elmegreen}}{2011}]{Elm11}
{Elmegreen} B.~G.,  2011, \mn@doi [\apj] {10.1088/0004-637X/737/1/10}, \href
  {https://ui.adsabs.harvard.edu/abs/2011ApJ...737...10E} {737, 10}

\bibitem[\protect\citeauthoryear{{F{\"o}rster Schreiber} et~al.,}{{F{\"o}rster
  Schreiber} et~al.}{2006}]{For06}
{F{\"o}rster Schreiber} N.~M.,  et~al., 2006, \mn@doi [\apj] {10.1086/504403},
  \href {https://ui.adsabs.harvard.edu/abs/2006ApJ...645.1062F} {645, 1062}

\bibitem[\protect\citeauthoryear{{Genkin} \& {Safronov}}{{Genkin} \&
  {Safronov}}{1975}]{Gen75}
{Genkin} I.~L.,  {Safronov} V.~S.,  1975, \sovast, \href
  {https://ui.adsabs.harvard.edu/abs/1975SvA....19..189G} {19, 189}

\bibitem[\protect\citeauthoryear{{Goldreich} \& {Lynden-Bell}}{{Goldreich} \&
  {Lynden-Bell}}{1965a}]{Gol65a}
{Goldreich} P.,  {Lynden-Bell} D.,  1965a, \mn@doi [\mnras]
  {10.1093/mnras/130.2.97}, \href
  {https://ui.adsabs.harvard.edu/abs/1965MNRAS.130...97G} {130, 97}

\bibitem[\protect\citeauthoryear{{Goldreich} \& {Lynden-Bell}}{{Goldreich} \&
  {Lynden-Bell}}{1965b}]{Gol65b}
{Goldreich} P.,  {Lynden-Bell} D.,  1965b, \mn@doi [\mnras]
  {10.1093/mnras/130.2.125}, \href
  {https://ui.adsabs.harvard.edu/abs/1965MNRAS.130..125G} {130, 125}

\bibitem[\protect\citeauthoryear{{Griv} \& {Gedalin}}{{Griv} \&
  {Gedalin}}{2012}]{Gri12}
{Griv} E.,  {Gedalin} M.,  2012, \mn@doi [\mnras]
  {10.1111/j.1365-2966.2012.20647.x}, \href
  {https://ui.adsabs.harvard.edu/abs/2012MNRAS.422..600G} {422, 600}

\bibitem[\protect\citeauthoryear{{Hunter}}{{Hunter}}{1972}]{Hun72}
{Hunter} C.,  1972, \mn@doi [Annual Review of Fluid Mechanics]
  {10.1146/annurev.fl.04.010172.001251}, \href
  {https://ui.adsabs.harvard.edu/abs/1972AnRFM...4..219H} {4, 219}

\bibitem[\protect\citeauthoryear{{Kim}, {Ostriker}  \& {Stone}}{{Kim}
  et~al.}{2002}]{Kim02}
{Kim} W.-T.,  {Ostriker} E.~C.,   {Stone} J.~M.,  2002, \mn@doi [\apj]
  {10.1086/344367}, \href
  {https://ui.adsabs.harvard.edu/abs/2002ApJ...581.1080K} {581, 1080}

\bibitem[\protect\citeauthoryear{{Kratter} \& {Lodato}}{{Kratter} \&
  {Lodato}}{2016}]{Kra16}
{Kratter} K.,  {Lodato} G.,  2016, \mn@doi [\araa]
  {10.1146/annurev-astro-081915-023307}, \href
  {https://ui.adsabs.harvard.edu/abs/2016ARA&A..54..271K} {54, 271}

\bibitem[\protect\citeauthoryear{{Law} et~al.,}{{Law} et~al.}{2022}]{Law22}
{Law} C.~J.,  et~al., 2022, \mn@doi [\apj] {10.3847/1538-4357/ac6c02}, \href
  {https://ui.adsabs.harvard.edu/abs/2022ApJ...932..114L} {932, 114}

\bibitem[\protect\citeauthoryear{{Lin} \& {Shu}}{{Lin} \& {Shu}}{1964}]{Lin64}
{Lin} C.~C.,  {Shu} F.~H.,  1964, \mn@doi [\apj] {10.1086/147955}, \href
  {https://ui.adsabs.harvard.edu/abs/1964ApJ...140..646L} {140, 646}

\bibitem[\protect\citeauthoryear{{Meidt}}{{Meidt}}{2022}]{Mei22}
{Meidt} S.~E.,  2022, arXiv e-prints, \href
  {https://ui.adsabs.harvard.edu/abs/2022arXiv220801888M} {p. arXiv:2208.01888}

\bibitem[\protect\citeauthoryear{{Nipoti} \& {Binney}}{{Nipoti} \&
  {Binney}}{2015}]{Nip15}
{Nipoti} C.,  {Binney} J.,  2015, \mn@doi [\mnras] {10.1093/mnras/stu2217},
  \href {https://ui.adsabs.harvard.edu/abs/2015MNRAS.446.1820N} {446, 1820}

\bibitem[\protect\citeauthoryear{{Rizzo}, {Vegetti}, {Fraternali}, {Stacey}  \&
  {Powell}}{{Rizzo} et~al.}{2021}]{Riz21}
{Rizzo} F.,  {Vegetti} S.,  {Fraternali} F.,  {Stacey} H.~R.,   {Powell} D.,
  2021, \mn@doi [\mnras] {10.1093/mnras/stab2295}, \href
  {https://ui.adsabs.harvard.edu/abs/2021MNRAS.507.3952R} {507, 3952}

\bibitem[\protect\citeauthoryear{{Romeo}}{{Romeo}}{1992}]{Rom92}
{Romeo} A.~B.,  1992, \mn@doi [\mnras] {10.1093/mnras/256.2.307}, \href
  {https://ui.adsabs.harvard.edu/abs/1992MNRAS.256..307R} {256, 307}

\bibitem[\protect\citeauthoryear{{Romeo} \& {Agertz}}{{Romeo} \&
  {Agertz}}{2014}]{Rom14}
{Romeo} A.~B.,  {Agertz} O.,  2014, \mn@doi [\mnras] {10.1093/mnras/stu954},
  \href {https://ui.adsabs.harvard.edu/abs/2014MNRAS.442.1230R} {442, 1230}

\bibitem[\protect\citeauthoryear{{Romeo} \& {Falstad}}{{Romeo} \&
  {Falstad}}{2013}]{Rom13}
{Romeo} A.~B.,  {Falstad} N.,  2013, \mn@doi [\mnras] {10.1093/mnras/stt809},
  \href {https://ui.adsabs.harvard.edu/abs/2013MNRAS.433.1389R} {433, 1389}

\bibitem[\protect\citeauthoryear{{Roychowdhury}, {Chengalur}, {Begum}  \&
  {Karachentsev}}{{Roychowdhury} et~al.}{2010}]{Roy10}
{Roychowdhury} S.,  {Chengalur} J.~N.,  {Begum} A.,   {Karachentsev} I.~D.,
  2010, \mn@doi [\mnras] {10.1111/j.1745-3933.2010.00835.x}, \href
  {https://ui.adsabs.harvard.edu/abs/2010MNRAS.404L..60R} {404, L60}

\bibitem[\protect\citeauthoryear{{Safronov}}{{Safronov}}{1960}]{Saf60}
{Safronov} V.~S.,  1960, Annales d'Astrophysique, \href
  {https://ui.adsabs.harvard.edu/abs/1960AnAp...23..979S} {23, 979}

\bibitem[\protect\citeauthoryear{{Shu}}{{Shu}}{1968}]{Shu68}
{Shu} F. H.-S.,  1968, PhD thesis, Harvard University, Massachusetts

\bibitem[\protect\citeauthoryear{{Spitzer}}{{Spitzer}}{1942}]{Spi42}
{Spitzer} Lyman J.,  1942, \mn@doi [\apj] {10.1086/144407}, \href
  {https://ui.adsabs.harvard.edu/abs/1942ApJ....95..329S} {95, 329}

\bibitem[\protect\citeauthoryear{{Tassoul}}{{Tassoul}}{1978}]{Tas78}
{Tassoul} J.-L.,  1978, {Theory of rotating stars}.
Princeton University Press

\bibitem[\protect\citeauthoryear{{Toomre}}{{Toomre}}{1964}]{Too64}
{Toomre} A.,  1964, \mn@doi [\apj] {10.1086/147861}, \href
  {https://ui.adsabs.harvard.edu/abs/1964ApJ...139.1217T} {139, 1217}

\bibitem[\protect\citeauthoryear{{Van Loo}, {Keto}  \& {Zhang}}{{Van Loo}
  et~al.}{2014}]{van14}
{Van Loo} S.,  {Keto} E.,   {Zhang} Q.,  2014, \mn@doi [\apj]
  {10.1088/0004-637X/789/1/37}, \href
  {https://ui.adsabs.harvard.edu/abs/2014ApJ...789...37V} {789, 37}

\bibitem[\protect\citeauthoryear{{Vandervoort}}{{Vandervoort}}{1970}]{Van70}
{Vandervoort} P.~O.,  1970, \mn@doi [\apj] {10.1086/150514}, \href
  {https://ui.adsabs.harvard.edu/abs/1970ApJ...161...87V} {161, 87}

\bibitem[\protect\citeauthoryear{{Wang}, {Klessen}, {Dullemond}, {van den
  Bosch}  \& {Fuchs}}{{Wang} et~al.}{2010}]{Wan10}
{Wang} H.-H.,  {Klessen} R.~S.,  {Dullemond} C.~P.,  {van den Bosch} F.~C.,
  {Fuchs} B.,  2010, \mn@doi [\mnras] {10.1111/j.1365-2966.2010.16942.x}, \href
  {https://ui.adsabs.harvard.edu/abs/2010MNRAS.407..705W} {407, 705}

\bibitem[\protect\citeauthoryear{{Yim}, {Wong}, {Xue}, {Rand}, {Rosolowsky},
  {van der Hulst}, {Benjamin}  \& {Murphy}}{{Yim} et~al.}{2014}]{Yim14}
{Yim} K.,  {Wong} T.,  {Xue} R.,  {Rand} R.~J.,  {Rosolowsky} E.,  {van der
  Hulst} J.~M.,  {Benjamin} R.,   {Murphy} E.~J.,  2014, \mn@doi [\aj]
  {10.1088/0004-6256/148/6/127}, \href
  {https://ui.adsabs.harvard.edu/abs/2014AJ....148..127Y} {148, 127}

\bibitem[\protect\citeauthoryear{{Yue}}{{Yue}}{1982}]{Yue82}
{Yue} Z.~Y.,  1982, \mn@doi [Geophysical and Astrophysical Fluid Dynamics]
  {10.1080/03091928208208998}, \href
  {https://ui.adsabs.harvard.edu/abs/1982GApFD..20....1Y} {20, 1}

\makeatother
\end{thebibliography}
\bibliographystyle{mnras}

\appendix

\section{Restrictions on the perturbation wavenumber}
\label{sec:app_local}

For the perturbation analysis to be consistent, the size of the
disturbance must be smaller than the characteristic length scales of
the unperturbed system. In particular, the properties of the
background must not vary significantly over the size of the
perturbations, so we must exclude from our analysis perturbations with
$k$ smaller than $1/\ell$, where $\ell\equiv |q|/||\grad q||$ is the
characteristic length over which any quantity $q$ varies in the
unperturbed configuration at the position of the disturbance.  An
estimate of $1/\ell^2$ is $||\grad \rho||||\grad
\press||/(\rho\press)$.  In general, in the presence of rotation, the
vertical density and pressure gradients are stronger than the
corresponding radial gradients, so we can take
$1/\ell^2\approx|\rhoz\pressz|/(\rho\press)$. Of course, the
underlying assumption is that the unperturbed gas distribution is
sufficiently smooth, so that $\ell$ can give a measure of macroscopic
gradients and is not affected by small-scale inhomogeneities.

As an additional restriction on the perturbation wavenumber, we also
require $k$ to be larger than $1/\L$, where $\L$ is the macroscopic
length scale of the gaseous system\footnote{$k>1/\ell$ does not
  necessarily imply $k>1/\L$: for instance $\ell\to \infty$ where
  $||\grad q||\to 0$.}.
In order to estimate $\L$, let us consider a very general argument
\citep[e.g.][]{Bin08,Ber14}: it follows from the virial theorem that
an equilibrium self-gravitating gaseous system of mass $M$ and sound
speed $\cs$ has characteristic size $\L\approx G M / \cs^2 $.  This
equation, combined with $M\approx \rho \L^3$, where $\rho$ is the mean
density of the system, gives $1/\L^2\approx G\rho/\cs^2\approx \kJ^2$,
where $\kJ^2=4\pi G\rho/\cs^2$ is the Jeans wavenumber squared.  So
the characteristic size of a self-gravitating gaseous system is of the
order of the Jeans length. This has the sometime overlooked
implication that, in the case of a gas cloud of finite size, the
classical Jeans stability analysis proves that linear perturbations
with $k\gtrsim \kJ$ are stable, but does not prove that modes with
$k\lesssim \kJ$ are unstable.  For a rotating flattened system the
shortest macroscopic scale is the vertical scale height, which is
typically of the order $1/\kJ$ (see e.g.\ the case of a vertical
isothermal distribution; Section~\ref{sec:case_study}).

The above simple arguments indicate that we must exclude from our
analysis modes with $k^2\lesssim |\rhoz\pressz|/(\rho\press)$ and
modes with $k\lesssim\kJ$. In practice, to approximately implement
both these conditions, we find it convenient to limit our analysis to
modes with $k$ satisfying \Eq(\ref{eq:k_restriction}).  This
restriction is adopted throughout Section~\ref{sec:crit}, with the
only exception of Section~\ref{sec:more_general_modes}.

\section{Analysis of the dispersion relations}
\label{sec:app_disp_rel}

In this Appendix we analyze dispersion relations in the form
$P(\omega,s)=0$, where $\omega$ is the frequency and $s\equiv\cs^2k^2$
with $k$ the wavenumber, which are biquadratic in $\omega$.  For given
$s$, we indicate the zeros of $P$ as $\omega^2_1$ and
$\omega^2_2>\omega^2_1$, and the discriminant of $P$ as
$\Deltaomega$. In the analysis, when dealing with a quadratic
polynomial of $s$, we indicate its discriminant as $\Deltas$ and its
zeros as $s_1$ and $s_2>s_1$.  When $\omega^2$ is real, the condition
for stability is $\omega^2_1>0$.  Modes such that $\omega^2$ is not
real are unstable (overstable), because there is at least one solution
with positive $\Im(\omega)$.

To simplify the notation we define the positive quantities $A\equiv
4\pi G\rho$, $B\equiv\kappa^2$, $C\equiv\nu^2$, $D\equiv \Nz^2$ and
$E\equiv \cs^2\hz^{-2}$, all with dimensions of a frequency squared,
as well as the dimensionless quantity $\zeta\equiv \kz^2/k^2$, which
is a measure of the relative contribution of the vertical component of
the wavevector.  The coefficients of the dispersion relations depend
in general on $A$, $B$, $C$, $D$, $E$ and $\zeta$. By definition
$0\leq\zeta\leq1$; we further assume $A>0$, $B>0$, $C>0$, $D>0$ and
$E>0$ (see Section~\ref{sec:crit}).  We recall that $C>D$ (see
\Eq\ref{eq:nznu}) and that in all the following sections, with the
only exception of Section~\ref{sec:disp_abcde}, we limit our analysis
to modes with $s>A+C$ (see \Eq\ref{eq:k_restriction}).

\subsection{Dispersion relations in the form 'ABCD$\zeta$'}
\label{sec:disp_abcdzeta}


We consider here dispersion relations in the form
\begin{equation}
  \omega^4+(A-B-C-s)\omega^2  +(1-\zeta)D s+\zeta B s -\zeta A B+BC=0.
  \label{eq:disp_abcdzeta}
\end{equation}
The discriminant of the dispersion relation is 
\begin{equation}
\begin{split}
  \Deltaomega&=
  (A-B-C-s)^2-4(1-\zeta)D s-4\zeta Bs+4\zeta AB-4BC\\
  &=s^2-2\left[A-B-C+2(1-\zeta)D+2\zeta B\right]s+(A-B-C)^2+4\zeta A B-4BC,
\end{split}
\end{equation}
which is positive for $s\to \infty$.  The discriminant of
$\Deltaomega(s)$ is
\begin{equation}
\begin{split}
  \Deltas&=
  4[A-B-C+2(1-\zeta)D+2\zeta B]^2-4(A-B-C)^2-16\zeta AB+16BC\\
  &=16(1-\zeta)\left[(A-B-C+D)D+BC-\zeta(B-D)^2\right].
\end{split}
\end{equation}
$\omega_1^2$ is given by
\begin{equation}
2\omega_1^2=s-(A-B-C)-\sqrt{\Deltaomega}.
\end{equation}
When $\Deltaomega>0$, given that $s>A+C>A-B-C$, the condition for
stability $\omega_1^2>0$ can be written as
\begin{equation}
s>\frac{A-\zeta^{-1}C}{1+\zeta^{-1}(1-\zeta)(D/B)},
\end{equation}
which is always satisfied. It follows that there is never monotonic instability.

When $(A-B-C+D)D+BC<0$, i.e.\
\begin{equation}
AD<(C-D)(D-B)\qquad\mbox{(sufficient for stability)}
\label{eq:suff_stab_abcdzeta}
\end{equation}  
$\Deltas<0$ (and thus $\Deltaomega>0$) $\forall \zeta$. Thus inequality~(\ref{eq:suff_stab_abcdzeta}) is a sufficient condition for
stability.

\subsection{Dispersion relations in the form 'ABC'}
\label{sec:disp_abc}

We consider here dispersion relations in the form
\begin{equation}
  \omega^4+(A-B-C-s)\omega^2+Bs-AB+BC=0.
    \label{eq:disp_abc}
\end{equation}
The discriminant of the dispersion relation is
\begin{equation}
  \Deltaomega=(A-B-C-s)^2-4Bs+4AB-4BC=
  [s-(A+B-C)]^2\geq 0,
\end{equation}
so $\omega^2$ is always real. For stability $\omega_1^2>0$
i.e.\ $s-(A-B-C)-\sqrt{[s-(A+B-C)]^2}>0$, which, for $s>A+C>A-B-C$
becomes $[s-(A-B-C)]^2>[s-(A+B-C)]^2>0$. If $s>(A+B-C)$ the latter
inequality reduces to $B>0$, which is always satisfied; if $s<A+B-C$
it reduces to $s>A+C$, which is always satisfied. Thus all modes are
stable.

\subsection{Dispersion relations in the form  'ABCDE'}
\label{sec:disp_abcde}


We consider here dispersion relations in the form
\begin{equation}
  \omega^4+\left(\frac{As}{s+E}-B-C-s\right)\omega^2+Ds+BC=0,
    \label{eq:disp_abcde}
\end{equation}
with $C>D$. Different from the rest of
Appendix~\ref{sec:app_disp_rel}, here we consider also modes with
$s<A+C$.  The discriminant of the dispersion relation is
\begin{equation}
  \Deltaomega=\left(A\frac{s}{s+E}-B-C-s\right)^2-4Ds-4BC,
\end{equation}
whose sign is determined by the sign of a third order polynomial in
$s$.  However, we can derive useful instability conditions even
without determining the sign of $\Deltaomega$.  When $\Deltaomega<0$
we have overstability.  When $\Deltaomega>0$, the condition to have
monotonic instability is $\omega_1^2<0$, i.e.\
\begin{equation}
-g(s)<\sqrt{g^2(s)-4Ds-4BC},
\end{equation}  
where
\begin{equation}
g(s)\equiv A\frac{s}{s+E}-B-C-s.
\end{equation}
The inequality is satisfied only when $g(s)>0$. Thus, if $g(s)>0$ and
$\Deltaomega>0$ we have monotonic instability. If $g(s)>0$ and
$\Deltaomega<0$ we have overstability. This implies that  a
sufficient condition for instability is $g(s)>0$, i.e.\
\begin{equation}
s^2-(A-B-C-E)s+E(B+C)<0\qquad\mbox{(sufficient for instability)},
\label{eq:two_crit_s}
\end{equation}
whose discriminant is $\Deltas=A^2-2(B+C+E)A+(B+C-E)^2$. The larger
root $s_2$ of the polynomial is given by
$2s_2=A-B-C-E+\sqrt{\Deltas}$. We have instability when $\Delta s>0$
and $s_2>0$. Imposing these two conditions we get
\begin{equation}
\frac{\sqrt{B+C}+\sqrt{E}}{\sqrt{A}}<1\qquad\mbox{(sufficient for instability)},
\label{eq:crit_rad_mod}
\end{equation}
which is thus a sufficient condition for instability.  We note that,
when combined with $s_2>0$, $\Delta s>0$ implies $s_1>0$: the interval
of unstable wavenumbers $s_1<s<s_2$ is centred at
$(A-B-C-E)/2>\sqrt{E(B+C)}$.

\subsection{Dispersion relations in the form  'ABCD'}
\label{sec:disp_abcd}


We consider here dispersion relations in the form
\begin{equation}
  \omega^4+(A-B-C-s)\omega^2+Ds+BC=0,
    \label{eq:disp_abcd}
\end{equation}
with $C>D$.
The discriminant of the dispersion relation is
\begin{equation}
  \Deltaomega=(A-B-C-s)^2-4Ds-4BC=s^2-2(A-B-C+2D)s+(A-B-C)^2-4BC.
\end{equation}
The discriminant of $\Deltaomega(s)$ is
\begin{equation}
  \Deltas=4(A-B-C+2D)^2-4(A-B-C)^2+16BC=16[(A-B-C+D)D+BC].
\end{equation}
When $\Deltaomega>0$, the condition for instability $\omega_1^2<0$
gives $Ds+BC<0$, which is never satisfied, so there is no monotonic
instability.  The conditions to have overstability are $\Deltas>0$ and
$s_2>A+C$. Given that $s_2=A-B-C+2D+s\sqrt{(A-B-C+2D)D+BC}$, these
conditions jointly lead to
\begin{equation}
AD> C(C-D)+(B/2)^2\qquad\mbox{(sufficient for instability)}.
\label{eq:suff_instab_abcd}
\end{equation}
When this condition is satisfied there are unstable (overstable) modes.

\section{Comparison with criteria for vertically stratified discs with polytropic equation of state}
\label{sec:app_hz}

In order to gauge the disc thickness parameter $\hz$ appearing in our
instability criterion~(\ref{eq:suff_instab_thick}) for vertically
stratified discs, here we compare our criterion with those found by
\citet{Gol65a} for uniformly rotating self-gravitating discs with
polytropic equation of state $\press\propto\rho^{\gammap}$,
considering in particular values of the polytropic index $\gammap=1$
and $\gammap=2$.  Our linear stability analysis, performed for
adiabatic perturbations, can be adapted to the case of a polytropic
equation of state simply imposing $\gamma=\gammap$, when the
unperturbed distribution is stratified with
$\press\propto\rho^\gammap$.  As a measure of the thickness $\hz$ we
can take the height $\hXperc$ of a strip centred on the midplane
containing a fraction $\xi=0.01X$ of the mass per unit surface:
$\hXperc=2\zxi$, with $\zxi$ such that
\begin{equation}
\frac{1}{\Sigma}\int_{-\zxi}^{z_\xi}\rho(z)\d z=\xi,
\end{equation}
where $\Sigma$ is given by \Eq(\ref{eq:surf_den}).  The following
analysis of the $\gammap=1$ (Section~\ref{sec:app_iso}) and
$\gammap=2$ (Section~\ref{sec:app_gam2}) cases suggests that a good
range of values of $\hz$ should be $\hsixty\lesssim \hz \lesssim
\heighty$.

\subsection{Self-gravitating isothermal disc with equation of state $\press\propto\rho$}
\label{sec:app_iso}

In the case of an isothermal disc, the vertical density distribution
is $\rho(z)=\rhozero\sech^2(z/b)$, where $\rhozero=\rho(0)$ and
$b=\csiso/\sqrt{2\pi G\rhozero}$ (\citealt{Spi42}; see also
Section~\ref{sec:case_study}), so
\begin{equation}
\hXperc=2b\atanh\xi.
\label{eq:hXperc_iso}
\end{equation}
\citet{Gol65a} found that a uniformly rotating ($\kappa^2=4\Omega^2$),
self-gravitating isothermal disc is unstable against isothermal
perturbations when\footnote{Recently, \citet{Mei22} claimed a lower
  threshold for instability $4\pi G\rhozero/\kappa^2>1$, i.e.\ $\pi
  G\rhobar/\kappa^2>1/6$, which is the limit of
  \Eq(\ref{eq:glb_like_crit_iso}) for $\hz\to \infty$. However, as far as
  we can tell, this threshold derives from including modes with
  $|\kz|<\kJ$, which should instead be excluded for consistency (see
  \Eq\ref{eq:k_restriction} and Appendix~\ref{sec:app_local})} $\pi
G\rhobar/\kappa^2>0.73$, where
\begin{equation}
  \rhobar\equiv\frac{1}{\Sigma}\int_{-\infty}^\infty\rho^2(z)\d
  z=\frac{2}{3}\rhozero
\end{equation}
is the mean gas density.  It is straightforward to show that for this
disc model, at given radius, the parameter $\Qthreed(z)$ defined in
\Eq(\ref{eq:suff_instab_thick}) attains its minimum at $z=0$, so a
sufficient condition to have instability at any height in the disc at
the considered radius is $\Qthreedmin=\Qthreed(0)<1$.  For this model,
imposing $\gamma=1$, the instability condition $\Qthreedmin<1$ can be
rewritten as
\begin{equation}
\frac{\pi G\rhobar}{\kappa^2}>\frac{1}{6}\left(1-\frac{1}{\sqrt{2}}\frac{b}{\hz}\right)^{-2}\qquad\mbox{(sufficient
  for instability)}.
\label{eq:glb_like_crit_iso}
\end{equation}
Using $\hz=\hXperc$ and \Eq(\ref{eq:hXperc_iso}), the
r.h.s.\ of the above equation equals 0.73 for $\xi\simeq0.59$, i.e.\ $\hz\approx \hsixty$.

\subsection{Self-gravitating polytropic disc with equation of state $\press\propto\rho^2$}
\label{sec:app_gam2}

In this case the vertical density distribution is given by \citep{Gol65a}
\begin{equation}
\rho(z)=\rhozero\cos\left(\frac{\pi}{2}\frac{z}{a}\right),
\label{eq:rho_slab_gam2}
\end{equation}  
for $|z|\leq a$ and $\rho=0$ for $|z|>a$, where $\rhozero=\rho(0)$ is
the density in the midplane, $a=\sqrt{\pi}\cszero/(4\sqrt{G\rhozero})$
is a characteristic scale length and
$\cszero=\sqrt{2\presszero/\rhozero}$ is the sound speed at $z=0$ with
$\presszero$ the pressure in the midplane. For this model
$\rhobar=(\pi/4)\rhozero$ and
\begin{equation}
\hXperc=\frac{4 a}{\pi}\arcsin\xi.
\label{eq:hXperc_gam2}
\end{equation}
\citet{Gol65a} found that this model is unstable against polytropic
$\gammap=2$ perturbations when $\pi G\rhobar/\kappa^2>1.11$.  As for
the isothermal disc (Section~\ref{sec:app_iso}), also in this case the
condition to have instability at any height at a given radius in the
disc is $\Qthreedmin=\Qthreed(0)<1$, which, imposing $\gamma=2$, can
be rewritten as
\begin{equation}
\frac{\pi G\rhobar}{\kappa^2}>\frac{\pi}{16}\left(1-\frac{2}{\pi}\frac{a}{\hz}\right)^{-2}\qquad\mbox{(sufficient
  for instability)}.
\label{eq:glb_like_crit_gam2}
\end{equation}
Using $\hz=\hXperc$ and \Eq(\ref{eq:hXperc_gam2}), the r.h.s.\ of the
above inequality equals 1.11 for $\xi\simeq0.76$, i.e.\ $\hz\approx
\heighty$.

\end{document}